\documentclass[%
 reprint,
 superscriptaddress,
 amsmath,amssymb,
 aps,
 floatfix,
]{revtex4-2}

\usepackage{capt-of}
\usepackage{float}
\usepackage{placeins}
\usepackage{graphicx}
\usepackage{siunitx}%
\usepackage{physics}
\usepackage{dcolumn}

\usepackage{epsfig}

\usepackage{multirow}

\usepackage{color}
\usepackage{orcidlink}
\usepackage{latexsym}
\usepackage{bm}
\usepackage{xcolor}
\usepackage[normalem]{ulem}
\usepackage{blkarray}

\begin{document}

\title{RIXS study of strain-tuned coupled valence and spin-state transition in epitaxial $(\text{Pr}_{0.85} \text{Y}_{0.15})_{0.7}\text{Ca}_{0.3}\text{CoO}_{3}$ films}

\author{Yiu-Fung Chiu\orcidlink{0000-0001-9673-6167}}
\email{yiufung.chiu@physics.ox.ac.uk}

\affiliation{Diamond Light Source, Harwell Campus, Didcot OX11 0DE, United Kingdom}

\affiliation{Department of Physics, University of Oxford, Clarendon Laboratory, Oxford OX1 3PU, United Kingdom}

\author{Jaewon Choi}
\altaffiliation[Present address: ]{Department of Physics \& Astronomy, Seoul National University, Seoul 08826, Republic of Korea}
\affiliation{Diamond Light Source, Harwell Campus, Didcot OX11 0DE, United Kingdom}

\author{Nileena Nandakumaran\orcidlink{0000-0003-3647-4008}}
\affiliation{Department of Chemical Engineering and Materials Science, University of Minnesota, Minneapolis, MN 55455, USA}

\author{Mirian Garcia-Fernandez\orcidlink{0000-0002-6982-9066}}
\affiliation{Diamond Light Source, Harwell Campus, Didcot OX11 0DE, United Kingdom}

\author{Jonathan  Pelliciari\orcidlink{0000-0003-1508-7746}}
\altaffiliation[Present address: ]{Department of Physics, Computer Science and Mathematics, University of Modena and Reggio Emilia, Modena, Italy}
\affiliation{National Synchrotron Light Source II, Brookhaven National Laboratory, Upton, New York 11973, USA}

\author{Shiyu Fan}
\affiliation{National Synchrotron Light Source II, Brookhaven National Laboratory, Upton, New York 11973, USA}

\author{Valentina Bisogni\orcidlink{0000-0002-7399-9930}}
\affiliation{National Synchrotron Light Source II, Brookhaven National Laboratory, Upton, New York 11973, USA}

\author{Chris Leighton\orcidlink{0000-0003-2492-0816}}
\affiliation{Department of Chemical Engineering and Materials Science, University of Minnesota, Minneapolis, MN 55455, USA}

\author{Ke-Jin Zhou\orcidlink{0000-0001-9293-0595}}
\affiliation{Diamond Light Source, Harwell Campus, Didcot OX11 0DE, United Kingdom}

\author{Stefano Agrestini\orcidlink{0000-0002-3625-880X}}
\email{stefano.agrestini@diamond.ac.uk}
\affiliation{Diamond Light Source, Harwell Campus, Didcot OX11 0DE, United Kingdom}

\date{\today}

\begin{abstract}

Strain-tuned $(\text{Pr}_{0.85} \text{Y}_{0.15})_{0.7}\text{Ca}_{0.3}\text{CoO}_{3}$ thin films provide a model system for controlling coupled valence, spin-state, structural and metal-insulator transitions (VSSTs) in perovskite cobaltite materials. Here, we use Co $L_{2,3}$ and Pr $M_{4,5}$ X-ray Absorption Spectroscopy (XAS), along with Co $L_{3}$ edge Resonant Inelastic X-ray Scattering (RIXS), to probe the local electronic structure across the strain-controlled VSST in epitaxial (Pr,Y,Ca)CoO$_{3}$ thin films. In compressively strained films on YAlO$_{3}$ (YAO) substrates, cooling through the transition produces Pr$^{4+}$-like spectral weight, suppresses Co$^{4+}$-like character, and favors predominantly low-spin Co$^{3+}$. The changes at the Pr and Co edges are quantitatively consistent with a reduction of approximately 0.15 holes per Co in the Co--O network. By contrast, tensile-strained thin films on La$_{0.18}$Sr$_{0.82}$Al$_{0.59}$Ta$_{0.41}$O$_{3}$ (LSAT) substrates have nearly temperature-independent spectra. Multiplet calculations reproduce the principal XAS lineshapes and local $dd$ excitations at high energy loss. Additionally, low-energy RIXS intensity recovered above the transition is consistent with a combination of Co$^{4+}$-like local excitations and a broader particle--hole continuum associated with the metallic phase. These results provide element-resolved evidence for charge compensation between the Pr and Co--O sublattices and support cooperative coupling among charge-transfer, spin-state, and lattice structure at the strain-controlled VSST.

\end{abstract}

\maketitle

\section{Introduction}

The rich variety of spin states and valence states of perovskite cobaltite compounds has been studied for decades, providing a versatile platform for exploring the interplay among orbital, spin, charge, and lattice degrees of freedom. In particular, the close competition among Co$^{3+}$ ($3d^{6}$) spin states in LaCoO$_3$ gives rise to a gradual, thermally driven spin-state crossover from a nonmagnetic low-spin ground state to a state with an increasing population of magnetic excited multiplets~\cite{Goodenough_1967,Haverkort_PRL,Podlesnyak}. Upon substituting La with Pr and Ca, this spin-state transition becomes first-order, and is coupled with a metal-insulator and structural transition~\cite{Tsubouchi_2002, Tsuboushi_2004}. It has also been shown that this transition is fundamentally linked to a valence transition of the Pr ions, from nominally Pr$^{3+}$ above the transition to a mixture of Pr$^{3+}$ and Pr$^{4+}$ below~\cite{Knizek, Herrero-Martin_2011, Herrero-Martin_2012,Garcia-Munoz_2011,Hejtmanek_EPB_2013}. Such valence transitions are possible in the case of Pr ions due to the unusual proximity of the Pr 4$f$ electrons to the Fermi level relative to other rare-earth ions~\cite{Chaturvedi}. In the half-doped Pr$_{0.5}$Ca$_{0.5}$CoO$_{3}$ compound, the transition temperature is at roughly 90~K. The bulk perovskite cobaltite $(\text{Pr}_{1-y} \text{Y}_{y})_{1-x}\text{Ca}_{x}\text{CoO}_{3}$ (PYCCO) has recently attracted considerable interest due to enhancement of the transition temperature of this coupled valence and spin-state transition (VSST) to 135~K, enabled by the Y ions stabilizing the low-temperature, low-volume phase~\cite{Hejtmanek_2010,Phelan_2014,Fujishiro_2012,Hejtmanek_EPB_2013}. The VSST occurs simultaneously with a metal-to-insulator (MIT) and a structural transition, suggesting strong correlation between these transitions. Much of the experimental work has focused on the characterisation of bulk samples of similar compounds, primarily involving Pr${}^{3+}$,  Pr$^{4+}$, Co$^{3+}$ high spin (HS, $S$ = 2) and low spin (LS, $S$ = 0), and  Co$^{4+}$ intermediate spin (IS, $S = 3/2$) states~\cite{Guillou}. A recent study on strain-tuned thin film samples of this compound family ($y = 0.15, x = 0.3$) has shown that the transition could be stabilized to near room temperature ($\sim$ 245~K) under compressive strain, while tensile strain suppresses this valence transition and induces a  long-range ferromagnetic metallic behavior, which is not seen in bulk~\cite{Chaturvedi,Dewey_2023,Dewey_2024}. Such control of the spin state and valence via strain shows great potential for devices, and motivates a microscopic investigation of how epitaxial strain controls the coupled valence and spin-state transition. In particular, the microscopic origin and causal sequence of the coupled transition remain incompletely understood.

Theoretical studies have mainly investigated the electronic aspect of the VSST by correlating the valence transition with the structural transition~\cite{Knizek,Chaturvedi}. In particular, Knížek $et~al.$ performed electronic-structure calculations within the generalized gradient approximation with an on-site Hubbard \(U\) correction (GGA+\(U\)), using experimentally determined crystal structures across the transition~\cite{Knizek}. These structures exhibit a contraction of the unit-cell volume and changes in the Co--O--Co bond angles, as observed by neutron diffraction~\cite{Baron_2010}. Their calculations showed that part of the occupied Pr 4$f$ states shifts above the Fermi level upon entering the low-temperature phase, transferring electrons to the hybridized Co(3$d$)--O(2$p$) states and giving rise to the valence transition. In this picture, the low-temperature valence transition is associated with a partial oxidation of Pr toward Pr$^{4+}$-like character and a compensating reduction of the Co--O network, thereby depleting Co$^{4+}$-like character across the structural transition and stabilizing the insulating state. This charge-transfer picture is supported by subsequent X-ray absorption and emission spectroscopy studies~\cite{Herrero-Martin_2011,Herrero-Martin_2012,Garcia-Munoz_2011,Hejtmanek_EPB_2013}. On warming, this charge transfer is reversed, restoring Co--O hole density and the metallic high-temperature phase. More recent DFT calculations have also shown that epitaxial strain modifies the relative stability of the competing metallic and insulating phases~\cite{Chaturvedi}. However, the fundamental cause of the structural and spin-state transition remains incompletely understood, leaving a gap for a complete microscopic mechanism of the VSST and how it is strain-tuned. Co $L$-edge Resonant Inelastic X-ray Scattering (RIXS) measurements could provide an ion-specific probe of the local spin-state and low-energy excitations that underlie the strain-controlled VSST, which is the motivation for the current work.

Previous work~\cite{Chaturvedi} established strain control of the VSST using thin film epitaxy, temperature dependent X-ray diffraction and electron energy loss spectroscopy, electronic transport, and magnetometry measurements. However, the local Co multiplet excitation spectrum across the strain-enhanced transition has not been directly resolved. This work provides the missing characterization of the local electronic states across the transition with synchrotron spectroscopic techniques by performing X-ray Absorption Spectroscopy (XAS) and RIXS experiments at the Co $L$ edge and Pr $M$ edge of two $(\text{Pr}_{0.85} \text{Y}_{0.15})_{0.7}\text{Ca}_{0.3}\text{CoO}_{3}$ samples under tensile and compressive strain on two distinct substrates. We report the ratio between different valence states and spin states across the transition, and use cluster multiplet calculation to extract the nature of low-energy excitations correlated with the transition. Finally, we discuss how the observed spin-state, charge-transfer, and lattice degrees of freedom may cooperate to stabilize the VSST under compressive strain and raise its transition temperature towards room temperature.

\begin{figure}
    \centering
    \includegraphics[width=0.7 \columnwidth]{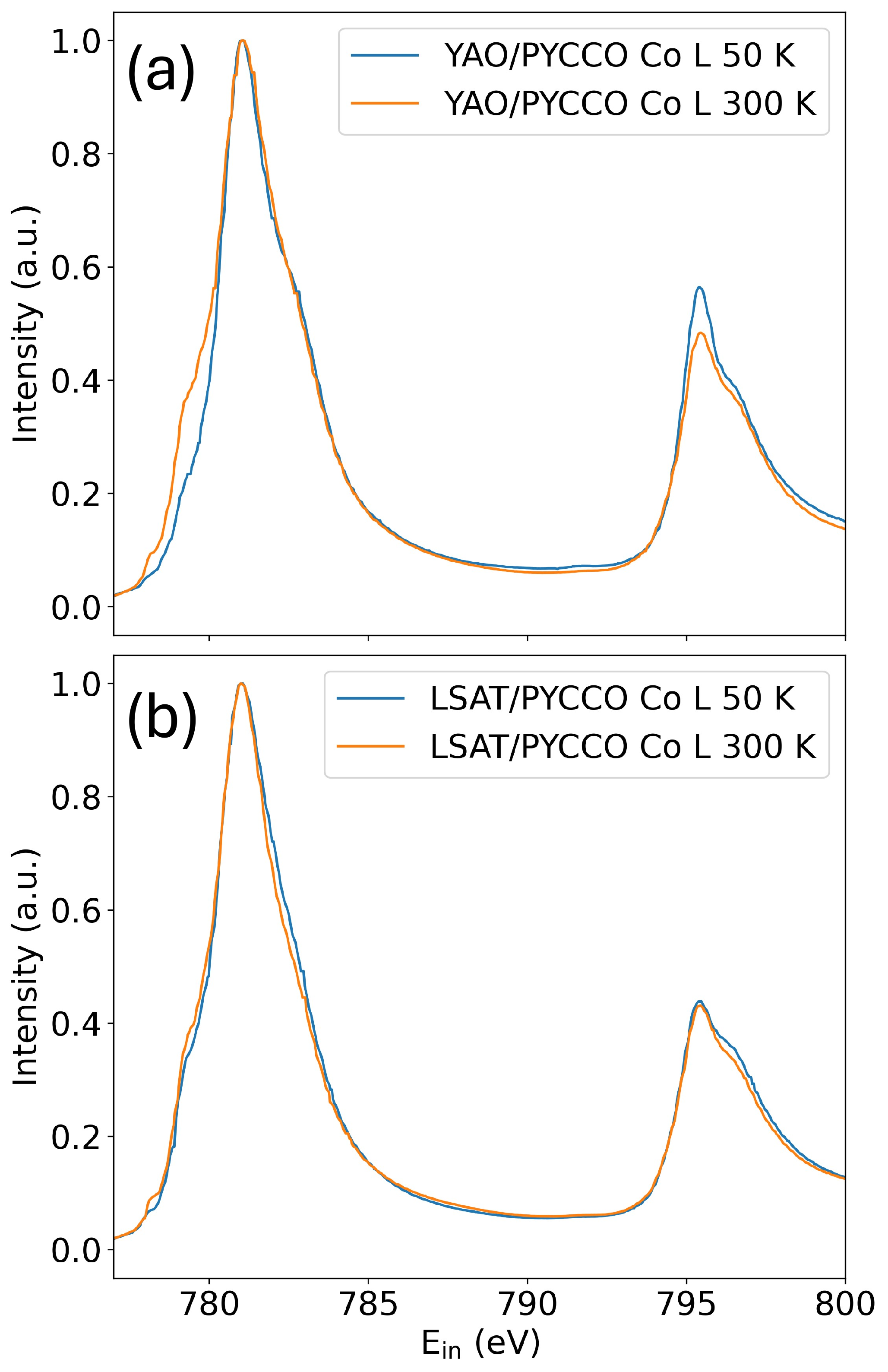}
    \caption{Co $L_{2,3}$-edge XAS data measured in TEY mode for (a) YAO/PYCCO below and above VSST and (b) LSAT/PYCCO at different temperatures.}
    \label{fig:1}
\end{figure}
\begin{figure}
    \centering
    \includegraphics[width= 0.7 \columnwidth]{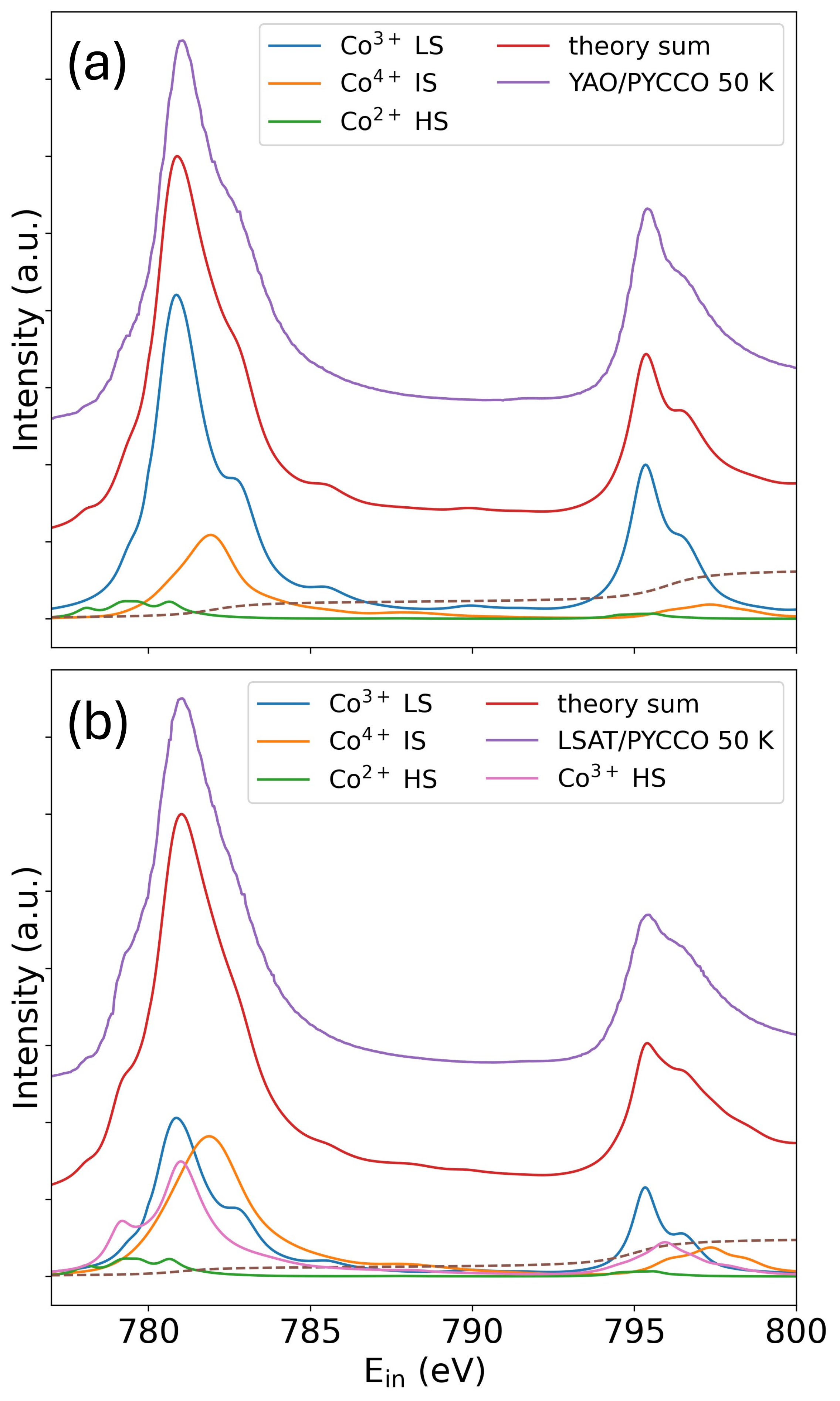}
    \caption{Experimental Co $L_{2,3}$-edge XAS spectra compared with population-weighted multiplet simulations for (a) YAO/PYCCO and (b) LSAT/PYCCO. The individual Co valence and spin-state components contributing to the simulated spectra are shown separately. Vertical offsets are applied to the experimental spectra and simulated sums for clarity.}
    \label{fig:2}
\end{figure}

\section{Experiment}

The synthesis of these epitaxial thin films was described in Refs.~\cite{Chaturvedi,Dewey_2023,Dewey_2024}. The thin films ($\sim$ 10.6 nm) were deposited onto substrates composed of YAlO${}_{3}$ (YAO) and  La$_{0.18}$Sr$_{0.82}$Al$_{0.59}$Ta$_{0.41}$O$_{3}$ (LSAT) to apply compressive (-2.1\%) and tensile heteroepitaxial strain (+2.34\%) respectively. Pr $M_{4,5}$ edge and Co $L_{2,3}$ edge XAS, and RIXS measurements were performed between 20~K and 300~K at the I21 beamline~\cite{Zhou2022I21} at the Diamond Light Source and SIX beamline at NSLS-II. The respective scattering angles were 154$^{\circ}$ and 150$^{\circ}$, with a horizontal scattering plane, a 20$^{\circ}$ incidence angle, and $\sigma$-polarised incident light. The corresponding RIXS energy resolutions were 31.5 and 30 meV. The data shown in the main figures were collected at I21.

\begin{figure}
    \centering
    \includegraphics[width= 0.8 \columnwidth]{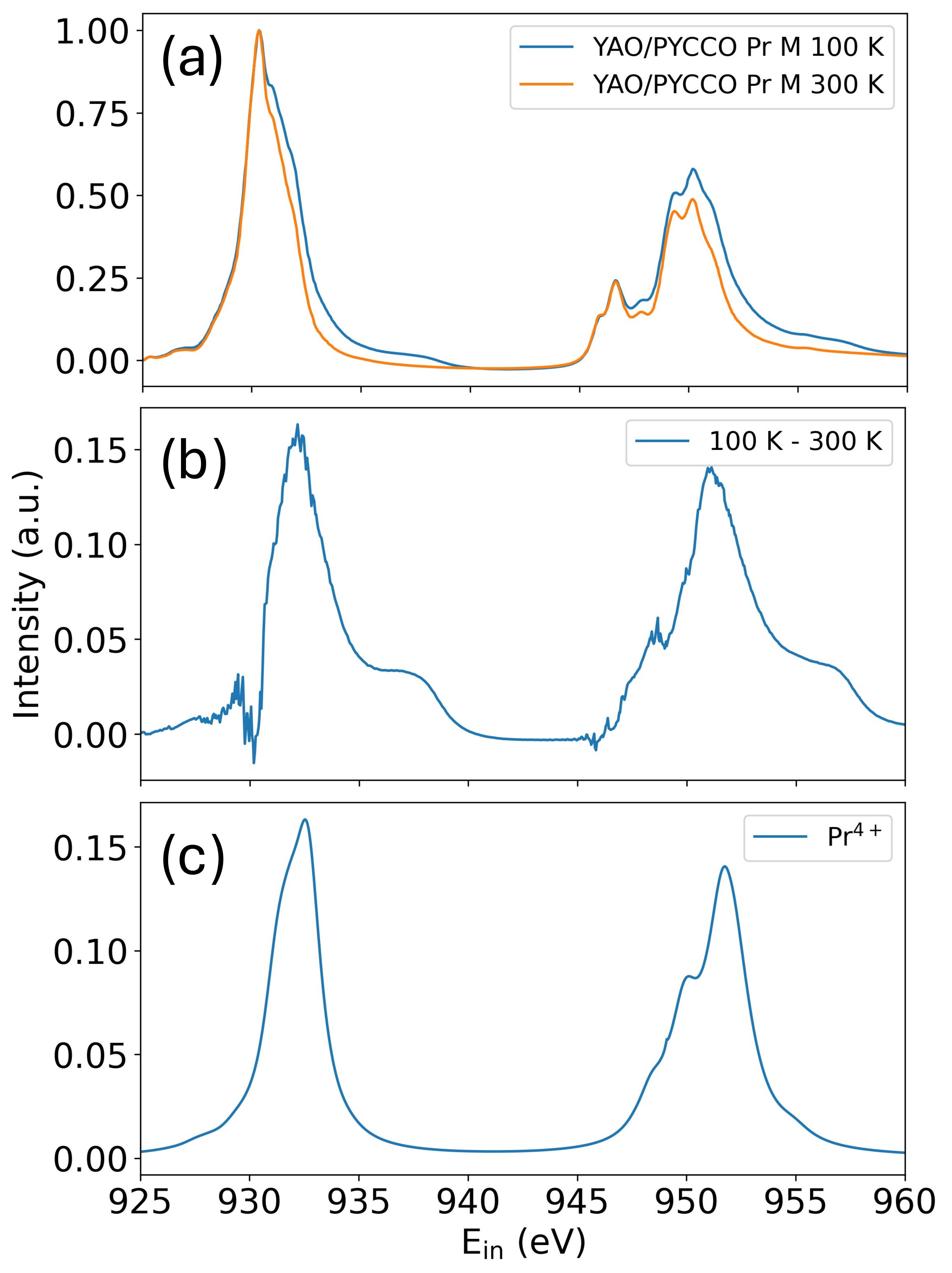}
    \caption{(a) Pr $M_{4,5}$-edge XAS TEY data for YAO/PYCCO below and above VSST. (b) XAS Difference spectrum between the two temperatures. (c) Simulation of Pr$^{4+}$ XAS spectra.}
    \label{fig:4}
\end{figure}

\begin{figure}
    \centering
    \includegraphics[width= 0.8 \columnwidth]{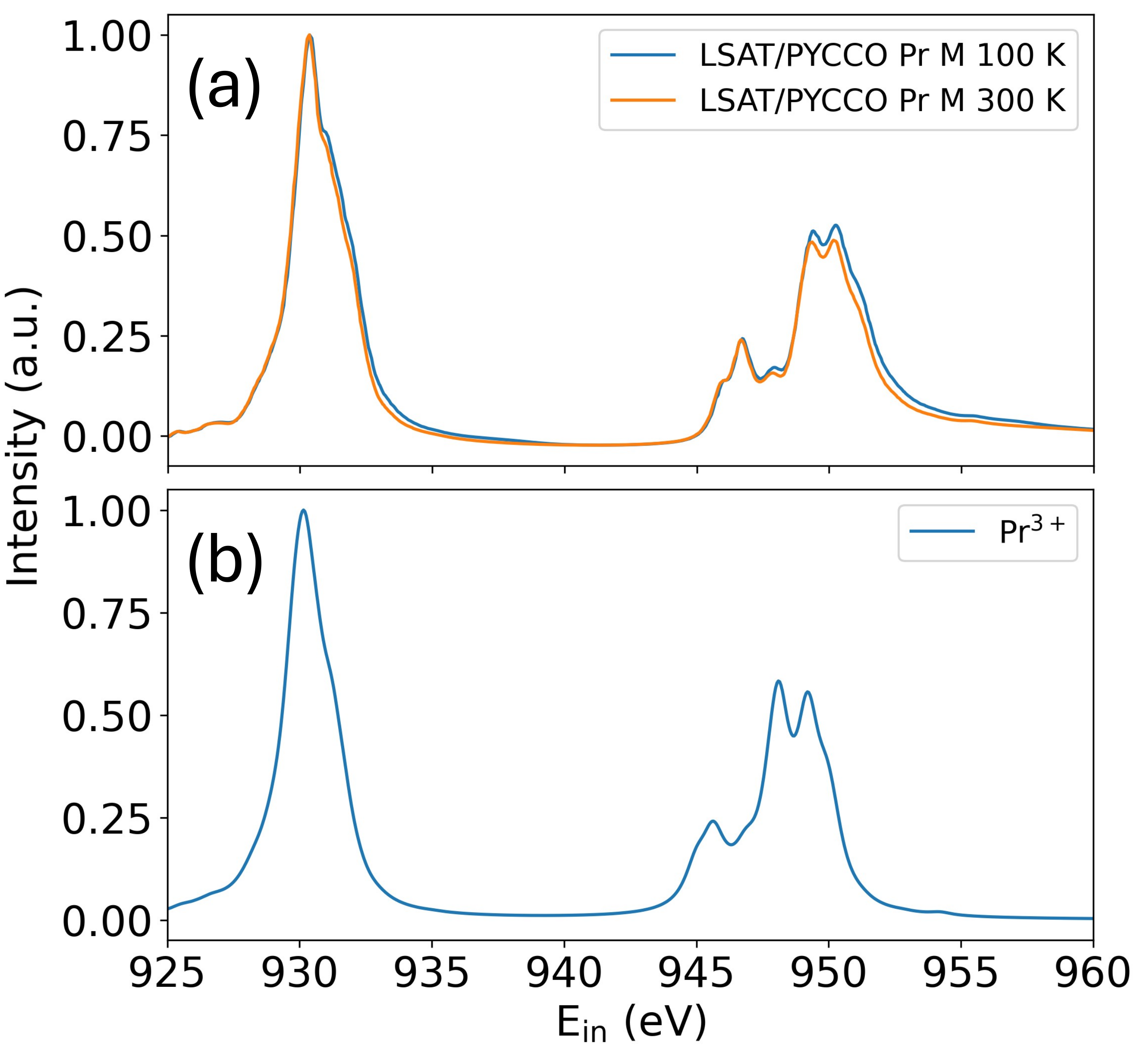}
    \caption{(a) Pr $M_{4,5}$-edge XAS TEY data for LSAT/PYCCO. (b) Simulation of Pr$^{3+}$ XAS spectra.}
    \label{fig:3}
\end{figure}

\section{Theory}

Multiplet ligand field theory calculations were performed using the Quanty code~\cite{Quanty}, which include the full atomic multiplet interactions, $2p$ and $3d$ spin--orbit coupling, octahedral crystal-field splitting, and Co $3d$--ligand hybridization. In the calculations we considered the
thermal population of the Co energy levels using a Boltzmann distribution. The Slater integrals were reduced to 70\% of their Hartree--Fock values for Co$^{2+}$ HS, Co$^{3+}$ HS and Co$^{4+}$ IS, and 75\% for Co$^{3+}$ LS. The parameters used in the calculations (in eV), listed in order $(10Dq,pd\sigma,pd\pi,\Delta,U_{dd},U_{pd})$, were Co$^{2+}$ HS: $(0.50,-1.55,0.72,6.50,6.50,8.00)$; Co$^{3+}$ HS: $(0.40,-1.55,0.72,2.00,5.50,7.00)$; Co$^{3+}$ LS: $(0.70,-1.70,0.74,2.00,5.50,7.00)$; and Co$^{4+}$ IS: $(0.15,-1.23,0.57,-5.00,5.50,7.00)$. To assess the competition between the Co$^{3+}$ spin states independently of the final line-shape refinements, we also varied $10Dq$ while keeping all other parameters fixed. The LS and HS configurations were found to be close in energy, with a modest change in $10Dq$ sufficient to reverse their energetic ordering.

\section{XAS results}

We first performed temperature-dependent XAS measurements on the YAO/PYCCO sample and the LSAT/PYCCO sample to investigate whether their electronic states exhibit the expected strain-dependent behaviour, with a VSST near 245~K in the compressively strained YAO/PYCCO film and no corresponding transition in the tensile LSAT/PYCCO film~\cite{Chaturvedi}. Fig.~\ref{fig:1}(a) shows the Co $L_{2,3}$ edge XAS data collected on the YAO/PYCCO sample, where the transition occurs at approximately 245~K. Both the total-electron-yield (TEY) and total-fluorescence-yield (TFY) (See Appendix A) on the YAO/PYCCO sample show a clear change in spectral lineshape across the VSST, indicating a strong change in the valence and spin states of the Co ions. For the PYCCO film on LSAT (Fig.~\ref{fig:1}(b)), however, the lineshape remains the same over the entire measured temperature range, which indicates that the valence and spin-state of the Co ions do not change, in agreement with the absence of the VSST indicated by prior spectroscopy, transport, and magnetometry measurements~\cite{Chaturvedi,Dewey_2023,Dewey_2024}. The TEY data show additional small peaks near 778~eV, indicative of Co$^{2+}$ ions, due to a surface-localized Co$^{2+}$-rich or CoO-like contribution on top of the thin film samples. These features are not observed in the more bulk-sensitive TFY data and thus should be negligible in the RIXS measurements.

We then used the XAS data to extract the ratio of different valence and spin states across the transition. The overall shape of the XAS closely resembles the related cobaltite compound $(\text{Pr}_{0.7} \text{Sm}_{0.3})_{0.7}\text{Ca}_{0.3}\text{CoO}_{3}$ in Ref.~\cite{Guillou}, so we start by assuming that below the transition, the sample is mainly composed of Co$^{3+}$ LS ions, the rest being Co$^{4+}$ IS. It should also be noted that it is still an open debate whether the Co$^{4+}$ is LS or IS~\cite{Guillou_2017}. The LS and IS Co$^{4+}$ configurations do not produce sufficiently distinct spectral signatures in the present data to allow an unambiguous assignment. We therefore refer to this contribution as a Co$^{4+}$-like component; the best fit was obtained by using the IS configuration. Above the transition, YAO/PYCCO exhibits spectral-weight fractions of Co$^{3+}$ LS, Co$^{3+}$ HS, and Co$^{4+}$-like components comparable to those observed in LSAT/PYCCO throughout the measured temperature range. We also added a small amount of Co$^{2+}$ HS ions to obtain a better fit to the experimental spectra. After refining the multiplet parameters using crystal field splittings obtained from RIXS energy maps and comparing with simulations, we obtain an estimate of the spectral-weight fraction in our samples, as listed in Table~\ref{tab:table}. Interestingly, above the VSST we find a slightly larger Co$^{3+}$ LS spectral-weight fraction compared to the bulk compound, but a similar ratio of total Co$^{3+}$ to Co$^{4+}$ ions for both samples. The measurements labeled as YAO are measured below the VSST unless specified otherwise, and all the measurements on YAO above the VSST are indistinguishable from the measurements on the LSAT/PYCCO samples within experimental uncertainty. The XAS simulation results according to this population are compared with experiment in Fig.~\ref{fig:2}. 

\begin{table}[H]
\centering
\vspace{2em}
\begin{tabular}{lcc}
\hline
Co ions\ \ \ \ \ \ \  & YAO (in \%) \ \ \ \  & LSAT (in \%) \\
        
\hline
Co$^{3+}$ LS & $\sim$ 80 \ \ & $\sim$ 40 \\
Co$^{3+}$ HS& 0 \ \  & $\sim$ 25 \\
Co$^{4+}$-like (IS)& $\sim$ 15 \ \  & $\sim$ 30 \\
\hline
\\
Co$^{2+}$ HS (TEY only)& $\sim$ 5 \ \  & $\sim$ 5 \\
\hline
\end{tabular}
\caption{Estimated Co spectral-weight fractions for PYCCO films on YAO and LSAT substrates at base temperature. Above the VSST, the films on YAO share the same spectral-weight fractions as the films on LSAT.}
\label{tab:table}
\end{table}

To find the change in the valence populations of the Pr ions, we compared the Pr $M$ edge XAS of YAO/PYCCO across the VSST as shown in Fig.~\ref{fig:4}(a). Additional spectral weight appears as high energy shoulders on the $M$$_{4}$ and $M$$_{5}$ main peaks. This additional spectral weight is due to the valence shift to a mixture of Pr$^{4+}$- and Pr$^{3+}$-like below the VSST. To quantify the valence population ratio, we used the difference spectrum shown in Fig.~\ref{fig:4}(b) to compare with a simulated pure Pr$^{4+}$ XAS spectrum (Fig.~\ref{fig:4}(c)) and adjusted the population ratio until the amplitude of the simulation matches the difference spectrum. As our main focus is on the amplitude of the XAS peaks, we have used an ionic model without any hybridization for the Pr XAS simulations. As a result, the broad features in the difference spectrum near
938 and 957~eV, which are consistent with Pr $4f$--O $2p$
hybridization effects reported previously
\cite{Herrero-Martin_2011,Butorin_1997}, are not reproduced by our ionic simulation. 

Importantly, we estimate that roughly 25\% of the Pr ions
develop Pr$^{4+}$-like character below the VSST. Given
that there are 0.595 Pr ions per formula unit, this
corresponds to approximately 0.15 electrons transferred
per Co ion from the Pr sublattice to the Co--O network.
Independently, the Co $L_{2,3}$-edge XAS analysis
indicates that the Co$^{4+}$-like fraction decreases from
approximately 30\% above the transition to 15\% below,
corresponding to a reduction of approximately 0.15
Co$^{4+}$-like holes per Co ion. The two independent estimates therefore agree in both sign and magnitude, providing quantitative support for charge compensation between the Pr and Co–O sublattices across the VSST, in agreement with previous studies of related bulk cobaltites
\cite{Guillou}. The equivalent measurements for the LSAT/PYCCO sample are shown in Fig.~\ref{fig:3}(a). The experimental data measured at 100~K and 300~K indicate no valence or spin-state change of the Pr ions, as for the Co ions. By a direct comparison with the Pr$^{3+}$ simulation in Fig.~\ref{fig:3}(b), we have identified that for the entire temperature range for LSAT/PYCCO and for YAO/PYCCO above the VSST, we have a population of Pr$^{3+}$ only, agreeing with the valence state population reported previously~\cite{Guillou,Garcia-Munoz_2011,Herrero-Martin_2011}. We have thus shown that the nature of the VSST in these strain-tuned PYCCO thin films is very similar to that of the relevant respective bulk samples.

\begin{figure}
    \centering
    \includegraphics[width= 1.1\columnwidth]{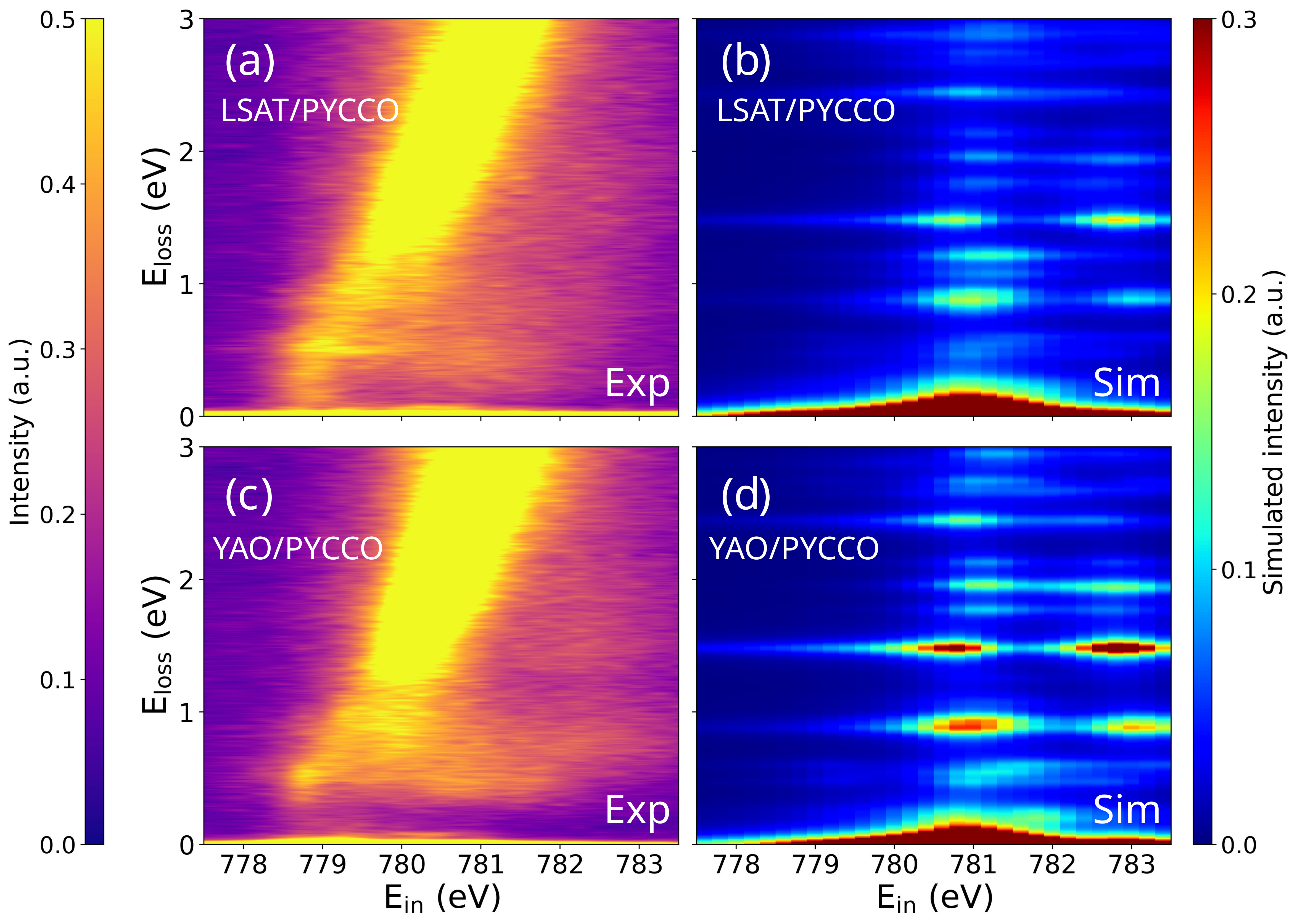}
    \caption{Experimental Co $L_{3}$-edge RIXS maps at 20~K compared with population-weighted multiplet simulations. (a, b) Experimental spectra and simulation for LSAT/PYCCO. (c, d) Experimental spectra and simulation spectra for YAO/PYCCO.}
    \label{fig:5}
\end{figure}

\begin{figure}
    \centering
    \includegraphics[width= 0.8\columnwidth]{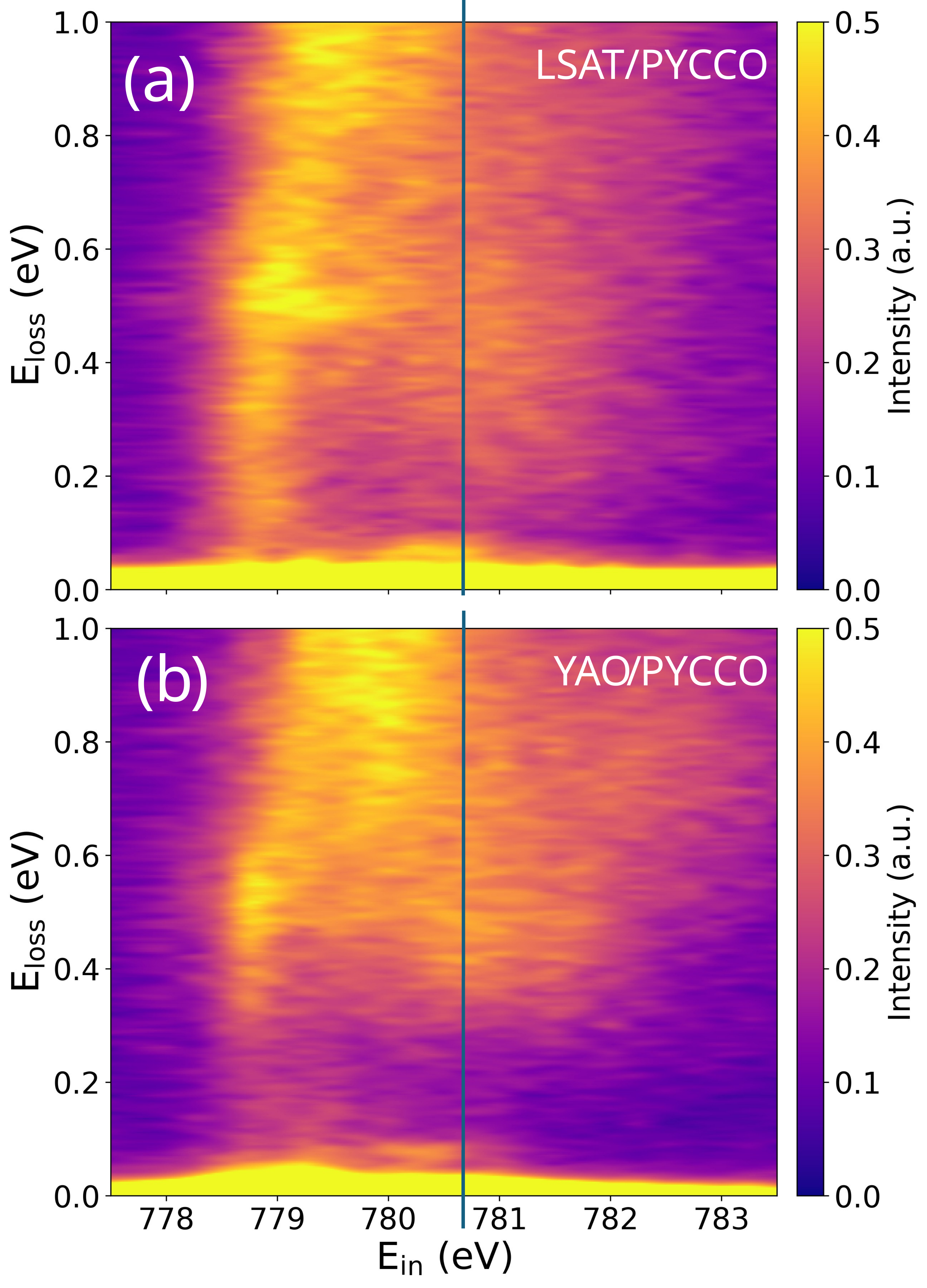}
    \caption{Co $L_{3}$-edge RIXS maps highlighting the low-energy excitations for (a) LSAT/PYCCO and (b) YAO/PYCCO. The vertical blue line marks the incident energy (780.6 eV) used for the spectra shown in Fig.~\ref{fig:7}.}
    \label{fig:6}
\end{figure}

\section{RIXS results}
We next use Co $L_{3}$-edge RIXS to identify the local excitations associated with the valence/spin-state reconstruction and extract the multiplet parameters across VSST. In Fig.~\ref{fig:5}, the experimental RIXS energy maps on YAO/PYCCO and LSAT/PYCCO samples at 20~K are presented alongside the population-weighted multiplet calculations based on the XAS-derived Co fractions, excluding the Co$^{2+}$ ions in the more bulk-sensitive RIXS. Both films show a broad fluorescence-like continuum, which obscures part of the high-energy multiplet structure, but $dd$ excitations at $\sim$ 0.5 eV and $\sim$ 0.9 eV remain visible and are reproduced in the multiplet calculations shown in Fig.~\ref{fig:5}(b,d). Thus, despite this fluorescence background, the reproducible $dd$ features provide constraints on the multiplet parameters. We note that the apparent intensity of the $dd$ excitations near 778.8~eV is enhanced by overlap with the broad fluorescence background, so they appear stronger there than in the simulations. According to the simulated spectra, there should be a strong 1.5~eV $dd$ excitation, which is not resolved experimentally. Its absence in the film spectra likely arises from the strong fluorescence background at this energy and deviations of the strained CoO$_{6}$ environment from the ideal cubic symmetry assumed in the multiplet calculation, which can split, broaden, and redistribute the calculated $dd$ spectral weight. A similar suppression or redistribution of high-energy \(d\)-\(d\) spectral weight has been reported in strained LaCoO\(_3\) films relative to the bulk~\cite{Wang_2019}.

A weak Raman-like feature near 80~meV is observed in both the YAO/PYCCO and LSAT/PYCCO spectra that resonates at 780.3~eV [see also Fig.~\ref{fig:6}]. Its energy lies close to that of the calculated $J=1\rightarrow J=3$ excitation within the Co$^{3+}$ HS manifold, making a local multiplet excitation a possible initial assignment. However, the feature is also present in the low-temperature YAO/PYCCO film, for which the XAS decomposition yields no detectable Co$^{3+}$ HS contribution and the ground state is nonmagnetic. We therefore assign this feature predominantly to a phonon excitation rather than to the Co$^{3+}$ HS multiplet. This interpretation is also consistent with high-energy oxygen phonons near 80~meV~\cite{Golosova} reported in related cobaltites. It is most clearly resolved at low temperature because of the narrower elastic and quasielastic response; at higher temperatures, it becomes obscured by the broadened elastic line, preventing a reliable determination of its temperature dependence.

\begin{figure}
    \centering
    \includegraphics[width= 1.1 \columnwidth]{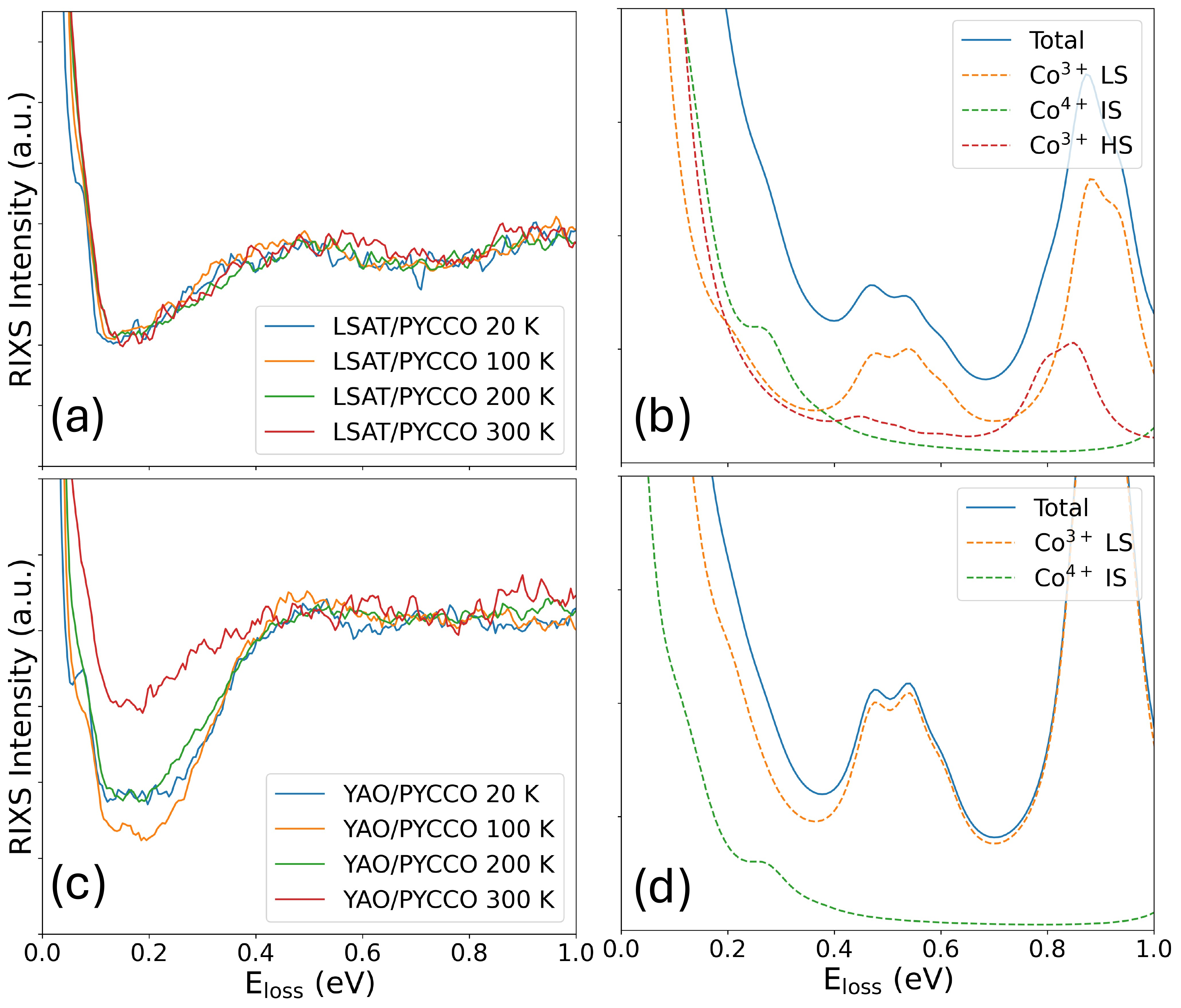}
    \caption{High-statistics temperature-dependent Co $L_{3}$-edge RIXS spectra compared to simulation. (a-b) Experimental spectra and simulation for LSAT/PYCCO. (c-d) Experimental spectra and simulation for YAO/PYCCO.}
    \label{fig:7}
\end{figure}

\begin{figure*}
    \centering
    \includegraphics[width= \textwidth]{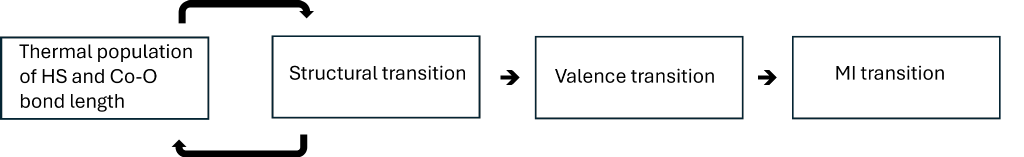}
    \caption{Qualitative schematic of the proposed cooperative mechanism underlying the VSST. The looped arrows indicate the possible positive feedback mechanism causing the first-order transition, which is coupled to the valence and metal-insulator transitions.}
    \label{fig:8}
\end{figure*}

The most robust RIXS signature of the VSST is instead found at low energy. As shown in Fig.~\ref{fig:6}, the low-temperature YAO/PYCCO film shows strongly suppressed spectral weight between $\sim$ 0.1 and 0.3~eV compared with the tensile LSAT/PYCCO film. This suppression is consistent with the insulating state, in which the available phase space for low-energy electronic excitations is reduced. To identify the nature of these spectral weights, we performed high-statistics temperature-dependent RIXS measurements at the energy along the vertical blue line indicated in the figure for both samples. The results are shown in Fig.~\ref{fig:7}, along with a cut of the simulated spectra for the base temperature for each sample. This spectral weight is recovered in the YAO/PYCCO case above the VSST and becomes essentially identical to that of the LSAT/PYCCO film, whereas the LSAT/PYCCO spectra remain nearly temperature independent. 

Within the present local multiplet model, part of this recovered low-energy intensity is reproduced by the Co$^{4+}$-like component, consistent with the increase of Co$^{4+}$-like/Co–O-hole character inferred independently from Co and Pr XAS~\cite{Guillou,Knizek,Herrero-Martin_2011,Herrero-Martin_2012,Garcia-Munoz_2011,Hejtmanek_EPB_2013}. However, the experimental low-energy spectral weight is not confined only to the Co$^{4+}$-like resonance condition. We therefore attribute the full redistribution to two related contributions: a local Co$^{4+}$-like multiplet component and a broader continuum-like electronic response associated with the metallic high-temperature phase. Such continuum-like RIXS intensity is expected when low-energy electron-hole excitations become available~\cite{Tomiyasu_2017}, and is not fully captured by a local cluster calculation~\cite{Hariki_2020}. The recovery of intensity toward the elastic line is therefore consistent with both the local Co valence/spin-state reconstruction and the filling or closing of the low-energy electronic gap across the metal-insulator transition.

\section{COOPERATIVE PICTURE OF THE VSST}

The controlled $10Dq$ calculations described above reveal close energetic competition between the Co$^{3+}$ LS and HS configurations. A modest change in the local crystal-field splitting is sufficient to switch their energetic ordering, supporting a cooperative feedback mechanism that could account for the strong enhancement of the VSST temperature under compressive heteroepitaxial strain. We start by assuming the system to be below the VSST, where the Co spectral weight consists primarily of
Co$^{3+}$ LS character, with a smaller Co$^{4+}$-like contribution modeled using the IS configuration. On warming, thermal expansion of the local Co--O environment may reduce the crystal-field splitting and favor increased HS Co$^{3+}$ character. The larger effective ionic radius~\cite{shannon1969effective} and increased $e_g$ occupation of HS Co$^{3+}$ could, in turn, promote further Co--O bond expansion, providing a positive spin--lattice feedback. This instability may couple to charge transfer between the Pr and Co--O sublattices and to the accompanying metal-insulator transition (see Fig.~\ref{fig:8}). However, the present data do not establish whether this spin-state redistribution precedes the structural transition or follows it. This provides a possible microscopic route by which the structural instability couples to the Pr-derived states near $E_F$ in DFT descriptions of the VSST~\cite{Knizek}, thereby linking the structural, valence, and metal-insulator transitions~\cite{Herrero-Martin_2011, Herrero-Martin_2012, Garcia-Munoz_2011, Hejtmanek_EPB_2013}. Importantly, this mechanism is specific to Pr-based cobaltites due to the unusual proximity of the occupied Pr $4f$ states to the Fermi level, whereas the occupied $4f$ states of other rare-earth ions, such as Nd, lie considerably deeper in energy and do not participate in comparable charge transfer~\cite{Knizek}.

Within this qualitative picture, compressive strain shifts the relative free energies of the competing phases toward the smaller-volume, Pr\(^{3+}\)/Pr\(^{4+}\)-mixed, LS Co\(^{3+}\)-rich insulating state. Warming can then favor the competing expanded phase through the combined effects of thermal expansion, spin-state entropy, increased HS Co\(^{3+}\) character, and recovery of Co–O hole density. Conversely, tensile strain stabilizes the expanded, HS Co\(^{3+}\) , Co\(^{4+}\)-like/Co–O-hole-rich metallic state over the measured temperature range, thereby suppressing the first-order transition into the collapsed low-spin phase. This picture is necessarily qualitative because the local CoO\(_6\) environment does not respond to biaxial strain in epitaxial films as it would to simple hydrostatic pressure; strain is instead accommodated through coupled changes in unit-cell volume, Pr–O bonding, octahedral tilts and distortions, and Co valence/spin state~\cite{Chaturvedi}.

\section{Conclusion}

We have performed XAS and RIXS on compressive and tensile strain-tuned thin films of $(\text{Pr}_{0.85} \text{Y}_{0.15})_{0.7}\text{Ca}_{0.3}\text{CoO}_{3}$ to estimate the Pr and Co valence populations and Co spin-state populations, and investigate their evolution with temperature. The compressively strained PYCCO film on YAO shows a clear redistribution of Pr valence, Co valence, and Co spin-state spectral weights across the near-room-temperature VSST at approximately 245~K, whereas the tensile-strained PYCCO film on LSAT remains largely temperature independent. The Pr M-edge spectra reveal an effective Pr$^{4+}$-like component below the transition, while Co $L_{2,3}$-edge XAS and Co $L_{3}$ RIXS indicate suppressed Co$^{4+}$ and dominant Co$^{3+}$ LS weight in the low-temperature insulating phase. The inferred redistribution of Co spectral weight is comparable to that reported for related bulk compounds. The $\sim$ 25\% Pr$^{4+}$-like component corresponds to roughly 0.15 electrons per Co transferred from the Pr sublattice to the Co–O network, consistent with the corresponding reduction of Co$^{4+}$-like hole character inferred from the Co $L_{2,3}$-edge analysis. Multiplet calculations reproduce the main Co $L_{2,3}$-edge XAS trends and identify the low-energy RIXS spectral weights recovered above the transition as associated with Co$^{4+}$ excitations and additional continuum-like low-energy electronic excitations in the metallic high-temperature phase. The RIXS response therefore reflects both the increase in the local Co$^{4+}$-like components and the increased low-energy particle–hole excitation phase space accompanying the metal-insulator transition across the VSST. The Pr and Co spectral changes are consistent with charge compensation between the Pr and Co–O sublattices, supporting the interpretation of the VSST in the strain-tuned thin films as a coupled Pr–Co charge-transfer process. The calculations further indicate close competition between Co$^{3+}$ LS and HS configurations, providing a natural route for strong spin-state–lattice coupling.

These results motivate a cooperative picture in which compressive strain stabilizes a collapsed, Pr$^{4+}$-admixed, Co$^{3+}$-rich LS insulating state, while warming restores HS Co$^{3+}$, and increases the population of Co$^{4+}$ and a predominant Pr$^{3+}$-like configuration. However, the present measurements do not uniquely determine whether the Co spin-state redistribution precedes or follows the structural transition. Future hysteresis-resolved RIXS/XAS, polarization-dependent RIXS maps and local structural probes of the CoO$_{6}$ environment will be required to establish the causal sequence of the strain-controlled VSST.


\begin{acknowledgments}
Y.F.C. acknowledges funding from Diamond Light Source (DLS) and the Clarendon Scholarship from the University of Oxford under joint doctoral studentship STU0477. Experiments were conducted at beamline I21 at DLS, and SIX at NSLS-II (proposal 314287). Work at the University of Minnesota (UMN) was supported by the Department of Energy (DOE) through the UMN Center for Quantum Materials under Grant No. DE-SC0016371. Parts of this work were conducted in the UMN Characterization Facility, which is partially supported by the National Science Foundation (NSF) through the MRSEC Program under Grant No. DMR-2011401. This research 
\newpage

\noindent
\begin{minipage}{\columnwidth}
    \centering
    \includegraphics[width=0.7\linewidth]{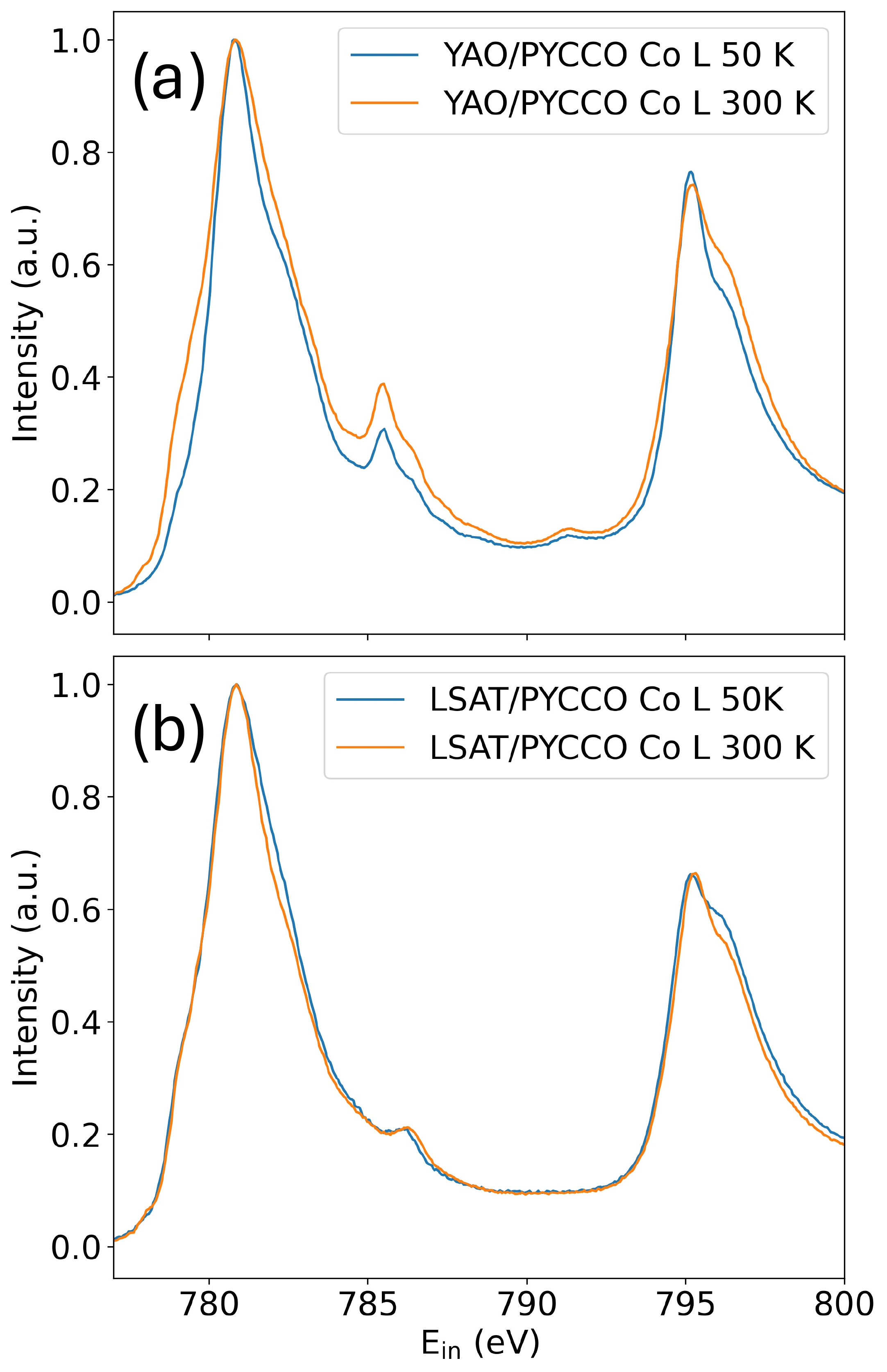}
    \captionof{figure}{Bulk-sensitive Co $L_{2,3}$-edge XAS spectra measured in TFY mode for (a) YAO/PYCCO below and above the VSST and (b) LSAT/PYCCO at different temperatures.}
    \label{fig:9}
\end{minipage}

\par\medskip
\noindent
used Beamline 2-ID of the National Synchrotron Light Source II, a U.S.~Department of Energy (DOE) Office of Science User Facility operated for the DOE Office of Science by Brookhaven National Laboratory under Contract No.~DE-SC0012704.  The work of S.F. and J.P. was supported by the Laboratory Directed Research and Development project of  Brookhaven National Laboratory No.~21-037 and 25-022.

\end{acknowledgments}

\appendix

\section{TFY XAS data}

To establish that the temperature-dependent spectral changes discussed in the main text are representative of the film bulk, we measured the Co $L_{2,3}$-edge XAS spectra in the more bulk-sensitive TFY mode, shown in Fig.~\ref{fig:9}. The YAO/PYCCO film exhibits a clear redistribution of spectral weights between 50 and 300~K, consistent with the VSST observed in TEY. In contrast, the TFY spectra of the LSAT/PYCCO film remain nearly temperature independent, consistent with the absence of this transition under tensile strain. Moreover, the weak Co$^{2+}$-like feature near 778~eV observed in TEY is absent in TFY, supporting its assignment to a surface CoO contribution. The additional feature near 786~eV in the TFY spectra is attributed to second-order incident photons producing Al $K$-edge absorption in the substrates and is therefore not intrinsic to the Co $L_{2,3}$-edge response.

\bibliography{biblio}

\end{document}